# Equivalence relations for Müller matrix symmetries of laboratory, LIDAR and planetary scattering geometries

Adrian J. Brown[*1]
[1] SETI Institute, 189 Bernardo Ave, Mountain View, CA 94043, USA

## Highlights

- Equivalence relations between the Müller matrix scattering geometries of laboratory, LIDAR and planetary observations are derived from first principles.
- Symmetries of the planetary and laboratory observational situations are established.
- Potential applications of these linkages are discussed and results compared to previous biomedical, atmospheric and planetary studies.

## Abstract

Symmetry relationships for optical observations of matter generally fall into several common scattering geometries. The "planetary" configuration is preferred among a group of observers of extraterrestrial planets, "laboratory" observations are performed in the biomedical research field and the LIDAR configuration is preferred among those using lasers to probe optical properties of horizontal surfaces with mirror or axial symmetry. This paper starts with the Stokes matrix formalism and uses symmetries of Müller matrix scattering to establishes links between the mathematical symmetries of each geometric configuration.

## Keywords

---

[*] Corresponding author address: Adrian J. Brown, SETI Institute, 189 Bernardo Ave, Mountain View, CA 94043; Tel: +1 (650) 810-0223; email: abrown@seti.org

## I. Introduction

Symmetry relationships are key to bringing insight and understanding to complicated optical situations. A survey of the modern polarization research literature quickly highlights that the Müller matrix polarization observations and experiments fall into several categories. One category is the "planetary" observers, who have been focused on observations of atmospheres around planetary bodies such as Venus or Jupiter [1–3]. On the other hand, the observational setup for "laboratory" (usually in the context of biomedical research [4–6]) or LIDAR observations [7–9] generally fall into another group. Although each group is peripherally aware of the other's work, there is as yet only limited cross comparison of results possible because of the different geometrical reference frames. In a recent publication, the need to establish links between these groups was emphasised [10].

Since most observational situations correspond to either 1.) laboratory - bistatic (source and detector widely separated) or 2.) planetary - collinear (source and detector parallel and aligned) polarization measurements, most polarization experiments will fall naturally into either of these two configurations. In this paper, the symmetries resulting form these configurations are established explicitly to further our understanding of backscattered electromagnetic radiation.

## II. Theory

We start with the Stokes vector, using it to describe an electric field, E [11,12] associated with a plane traveling wave with perpendicular and parallel amplitudes $E_\perp$ and $E_\parallel$:

$$\begin{aligned}
I &= \langle E_\parallel E_\parallel^* \rangle + \langle E_\perp E_\perp^* \rangle \\
Q &= \langle E_\parallel E_\parallel^* \rangle - \langle E_\perp E_\perp^* \rangle \\
U &= \langle E_\parallel E_\perp^* \rangle + \langle E_\perp E_\parallel^* \rangle \\
V &= i(\langle E_\parallel E_\parallel^* \rangle + \langle E_\perp E_\perp^* \rangle)
\end{aligned} \quad (1)$$

where angle brackets $\langle \rangle$ indicate a time average and $^*$ indicates complex conjugation.

The Müller matrix of an optical system is represented by elements $M_{11} \ldots M_{44}$. A generic optical system may then be represented by the system's Müller matrix $\boldsymbol{M}$, times the input Stokes vector, $I_0$:

$$\boldsymbol{I} = \boldsymbol{M} I_0 = \begin{pmatrix} M_{11} & M_{12} & M_{13} & M_{14} \\ M_{21} & M_{22} & M_{23} & M_{24} \\ M_{31} & M_{32} & M_{33} & M_{34} \\ M_{41} & M_{42} & M_{43} & M_{44} \end{pmatrix} \begin{pmatrix} I_0 \\ Q_0 \\ U_0 \\ V_0 \end{pmatrix} \quad (2)$$

For any *scattering matrix* $F(\mu,\theta)$, temporarily suppressing the $\mu, \theta$ dependence, the law of reciprocity [13,14] requires that:

$$F = \Delta_3 F^T \Delta_3 \tag{3}$$

where $^T$ indicates transpose and $\Delta_3$ is defined as:

$$\Delta_3 = \begin{pmatrix} 1 & 0 & 0 & 0 \\ 0 & 1 & 0 & 0 \\ 0 & 0 & -1 & 0 \\ 0 & 0 & 0 & 1 \end{pmatrix} \tag{4}$$

A similar relation exists for media with isotropic mirror symmetry. The mirror symmetry relation implies that

$$F = \Delta_{3,4} F \Delta_{3,4} \tag{5}$$

where $\Delta_{3,4}$ is defined as:

$$\Delta_{3,4} = \begin{pmatrix} 1 & 0 & 0 & 0 \\ 0 & 1 & 0 & 0 \\ 0 & 0 & -1 & 0 \\ 0 & 0 & 0 & -1 \end{pmatrix} \tag{6}$$

For any target for which the mirror symmetry and law of reciprocity holds, the scattering matrix is reduced to 6 independent elements [12] as follows:

$$F = \begin{pmatrix} a_1 & b_1 & 0 & 0 \\ b_1 & a_2 & 0 & 0 \\ 0 & 0 & a_3 & b_2 \\ 0 & 0 & -b_2 & a_4 \end{pmatrix} \tag{7}$$

The rotation matrix for an optical system for counterclockwise rotation (looking in the direction of propagation) by an angle $\alpha$ is defined as:

$$L(\alpha) = \begin{pmatrix} 1 & 0 & 0 & 0 \\ 0 & \cos 2\alpha & \sin 2\alpha & 0 \\ 0 & -\sin 2\alpha & \cos 2\alpha & 0 \\ 0 & 0 & 0 & 1 \end{pmatrix} \tag{8}$$

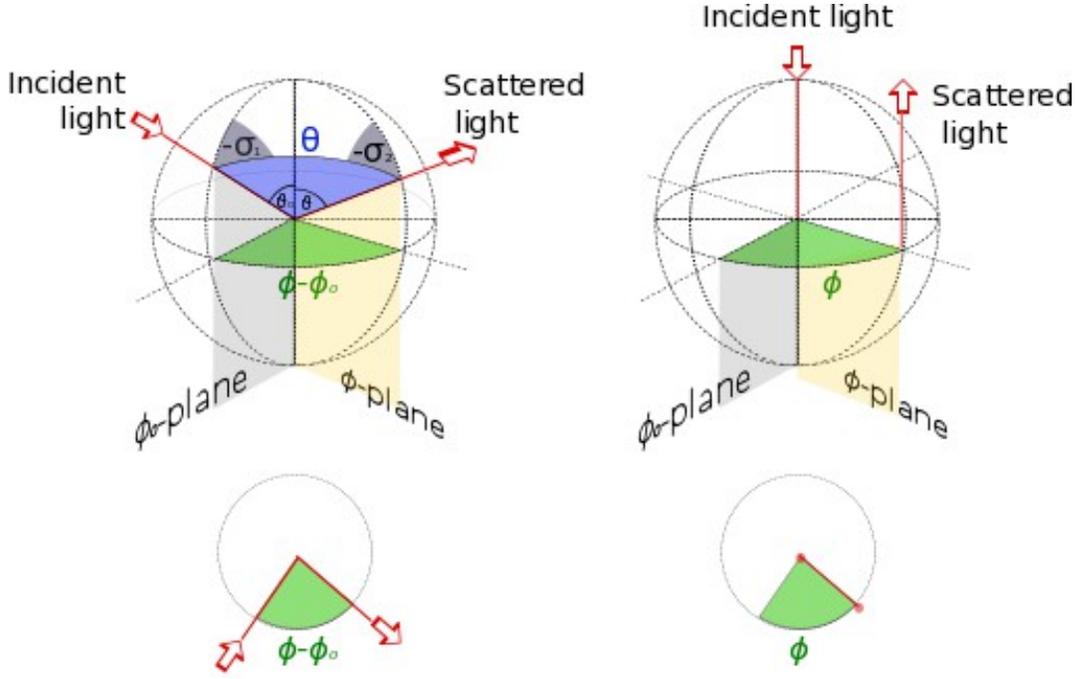

Figure 1a. (left) Planetary Observation Geometry 1b. (right) Laboratory or LIDAR Observation Geometry

*Planetary observations.* The planetary scattering geometry is represented in Fig 1a. Typically this is a bistatic measurement where source and detector are separated by a large distance. When an observation is made of a planetary surface or atmosphere, for off-zenith illumination, two rotations of the plane of reference are required to transform the incident beam into the scattered beam in the following manner:

$$Z_{planet}(\theta, \sigma_1, \sigma_2) = L(\pi - \sigma_2) F(\theta) L(-\sigma_1)$$
$$Z_{planet}(\theta, \sigma_1, \sigma_2) = L(-\sigma_2) F(\theta) L(-\sigma_1) \tag{9}$$

The second relation comes from the periodic characteristics of *L*. *Z* is termed the *phase matrix [15]*. Expanding this equation using (7) and (8), we get:

$$Z_{planet}(\theta, \sigma_1, \sigma_2) = \begin{vmatrix} a_1 & b_1 C_1 & -b_1 S_1 & 0 \\ b_1 C_2 & a_2 C_1 C_2 - a_3 S_1 S_2 & -a_2 S_1 C_2 - a_3 C_1 S_2 & -b_2 S_2 \\ b_1 S_2 & a_2 C_1 S_2 + a_3 C_2 S_1 & -a_2 S_1 S_2 + a_3 C_1 C_2 & b_2 C_2 \\ 0 & -b_2 S_1 & -b_2 C_1 & a_4 \end{vmatrix} \tag{10}$$

where $C_1 = \cos 2\sigma_1$, $C_2 = \cos 2\sigma_2$, $S_1 = \sin 2\sigma_1$, $S_2 = \sin 2\sigma_2$

As discussed in Brown and Xie [10], a special geometry exists for observational geometries where $\sigma_2 = 0$ (observer at zenith, source anywhere). In this case, the following equation holds:

$$Z_{planet}(\theta,\sigma_1,0)=\begin{vmatrix} a_1 & b_1 C_1 & -b_1 S_1 & 0 \\ b_1 & a_2 C_1 & -a_2 S_1 & 0 \\ 0 & a_3 S_1 & a_3 C_1 & b_2 \\ 0 & -b_2 S_1 & -b_2 C_1 & a_4 \end{vmatrix} \qquad (11)$$

When σ₁=0 (observer anywhere, source at zenith), the following relationship holds for the planetary geometry:

$$Z_{planet}(\theta,0,\sigma_2)=\begin{vmatrix} a_1 & b_1 & 0 & 0 \\ b_1 C_2 & a_2 C_2 & -a_3 S_2 & -b_2 S_2 \\ b_1 S_2 & a_2 S_2 & a_3 C_2 & b_2 C_2 \\ 0 & 0 & -b_2 & a_4 \end{vmatrix} \qquad (12)$$

*Laboratory or LIDAR observations.* In a laboratory or LIDAR observation, the source and detector are typically collinear, as seen in Fig. 1b. When an observation is made in the laboratory or by a LIDAR of a horizontal surface, just two rotations of the observation frame are necessary [16], however, each rotation is of the same angle (in opposite directions), as follows:

$$\begin{aligned} Z_{lab}(\theta,\sigma) &= L(\pi-\sigma)F(\theta)L(\sigma) \\ Z_{lab}(\theta,\sigma) &= L(-\sigma)F(\theta)L(\sigma) \end{aligned} \qquad (13)$$

Note that relation (13) is similar in theme, but different in detail to those proposed by various previous workers[7,8,16–18]. It is most similar to equation (13) of [8], although it was not used in the same manner by those authors. As will be discussed further below, this formulation is the only one that can match existing experimental results in the literature [5]. Again, the second relation comes from the periodic characteristics of L. Expanding (13) using (7) and (8) or putting $C=C_1=C_2$ and $S=S_1=S_2$ in (10) one obtains:

$$Z_{lab}(\theta,\sigma)=\begin{vmatrix} a_1 & b_1 C & b_1 S & 0 \\ b_1 C & a_2 C^2+a_3 S^2 & (a_2-a_3)CS & -b_2 S \\ b_1 S & (a_2-a_3)CS & a_2 S^2+a_3 C^2 & b_2 C \\ 0 & b_2 S & -b_2 C & a_4 \end{vmatrix} \qquad (14)$$

As for the planetary observations, special geometries exist in the laboratory situation when σ=0. For this case, the following scattering matrix holds:

$$Z_{lab}(\theta,0)=\begin{vmatrix} a_1 & b_1 & 0 & 0 \\ b_1 & a_2 & 0 & 0 \\ 0 & 0 & a_3 & b_2 \\ 0 & 0 & -b_2 & a_4 \end{vmatrix} \qquad (15)$$

*Comparisons of the rotational symmetries of configurational states*. The differences and similarities between the 'Planetary' and 'Laboratory/LIDAR' modes are thus summarized in equations (10) and (14). In order to break the degeneracy of equation (10) we make a choice to view from the zenith with $\sigma_1=0$, thus obtaining (12).

The differences in rotational symmetries of the two configurations can be characterized in the following manner. Setting the scattering coefficients $a_1$, $a_2$, $a_4$, $b_1$ and $b_2=1$ and $a_3=0.5$, we can evaluate the trigonometric dependence on the scattering geometry. Carrying out this substitution in equations (12) and (14), we arrive at the results in Figure 2.

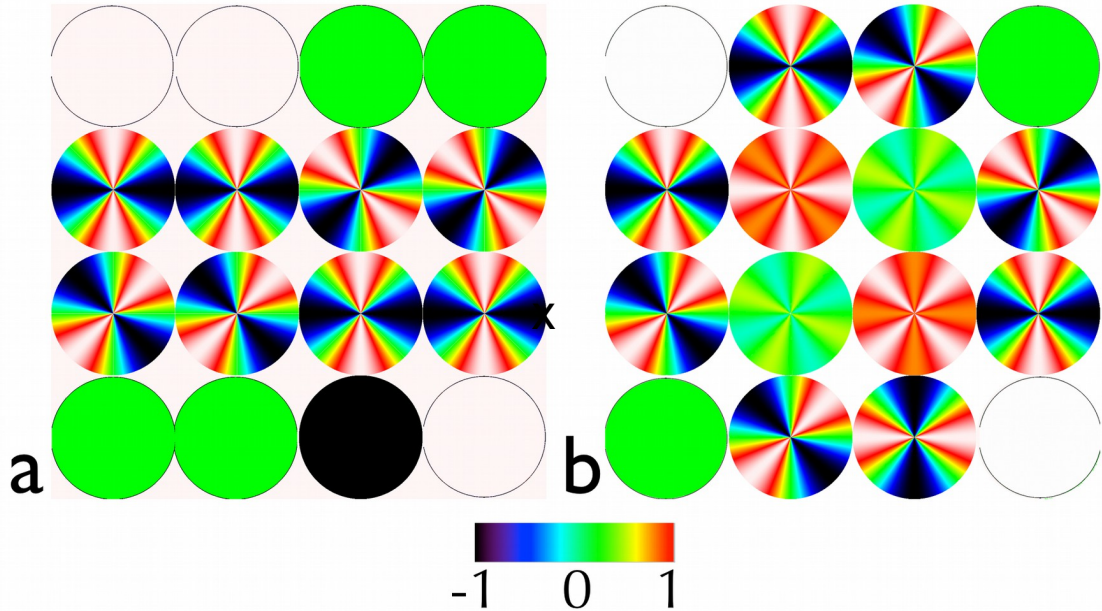

Figure 2 - Rotational Symmetries of (a) planetary (eqn 12) and (b) laboratory (eqn 14) observations. Note that to match the standard representations of Müller matrix hemispherical maps, the y-axis points down in these images. All coefficients except $a_3$ have been set to 1 to emphasize the rotational dependences. See text for further discussion.

By comparisons of any Müller matrix hemispherical maps with Figure 2, we can quickly establish the nature of the geometric configuration that was used to produce the map. As a useful mnemonic, the laboratory configuration is more 'complicated' because the there are more constant matrix elements (8 versus 4) and the central matrix elements ($M_{22}$, $M_{23}$, $M_{33}$, $M_{32}$) have 2-fold rotational symmetry rather than 4-fold rotational symmetry of their planetary counterparts.

*Equivalence relations*. The equivalence relations for the zenith laboratory (eqn (12) ) and planetary (eqn (14) ) observations can therefore be expressed as:

$$a_{1\text{planet}} = a_{1\text{lab}}$$
$$a_{2\text{planet}} = \frac{C^2}{C_2} a_{2\text{lab}} + \frac{S^2}{C_2} a_{3\text{lab}}$$
$$a_{3\text{planet}} = \frac{S^2}{C_2} a_{2\text{lab}} + \frac{C^2}{C_2} a_{3\text{lab}}$$
$$a_{4\text{planet}} = a_{4\text{lab}} \tag{16}$$
$$b_{1\text{planet}} = b_{1\text{lab}} \cdot \frac{S}{S_1}$$
$$b_{2\text{planet}} = b_{2\text{lab}} \cdot \frac{S}{S_1}$$

All of these relations can be evaluated from experimental measurements of the same scattering target in either frame, and then these may be used to convert to the other scattering geometry.

### III. Analytical Example

To test the veracity of this formulation outlined above, we present a simple application of the ideas presented here, using the analytically tractable electric dipole or Rayleigh scattering matrix [19] as an example.

The scattering matrix for a dipole as discussed in [20] is given by:

$$F_{dipole}(\theta) = \begin{pmatrix} (3/4)(1+\cos^2(\theta)) & (-3/4)(\sin^2(\theta)) & 0 & 0 \\ (-3/4)(\sin^2(\theta)) & (3/4)(1+\cos^2(\theta)) & 0 & 0 \\ 0 & 0 & (3/2)\cos(\theta) & 0 \\ 0 & 0 & 0 & (3/2)\cos(\theta) \end{pmatrix} \tag{17}$$

When we use this scattering matrix in equation (9) for the planetary scattering geometry, we get:

$$Z_{planet}(\theta, \sigma_1, \sigma_2) = \begin{pmatrix} (3/4)(1+c^2) & (-3/4)C_1 s^2 & (3/4)S_1 s^2 & 0 \\ (-3/4)C_2 s^2 & (3/4)(1+c^2)C_1 C_2 - (3/2)c S_1 S_2 & (-3/4)(1+c^2)S_1 C_2 - (3/2)c C_1 S_2 & 0 \\ (-3/4)S_2 s^2 & (3/4)(1+c^2)C_1 S_2 + (3/2)c S_1 C_2 & -(3/4)(1+c^2)S_1 S_2 + (3/2)c C_1 C_2 & 0 \\ 0 & 0 & 0 & (3/2)c \end{pmatrix} \tag{18}$$

where c = cos$\theta$, s = sin$\theta$, $C_1$ = cos2$\sigma_1$, $S_1$ = sin2$\sigma_1$, $C_2$ = cos2$\sigma_2$, $S_2$ = sin2$\sigma_2$
For the special case where $\sigma_1$=0 (normal incident light), we have the following:

$$Z_{planet}(\theta,0,\sigma_2) = \begin{vmatrix} (3/4)(1+c^2) & (-3/4)s^2 & 0 & 0 \\ (-3/4)C_2 s^2 & (3/4)(1+c^2)C_2 & (3/2)c S_2 & 0 \\ (-3/4)S_2 s^2 & (3/4)(1+c^2)S_2 & (3/2)c C_2 & 0 \\ 0 & 0 & 0 & (3/2)c \end{vmatrix} \quad (19)$$

For the laboratory scattering geometry, we substitute equation (8) and (17) into equation (13), to get:

$$Z_{lab}(\theta,\sigma,\sigma) = \begin{vmatrix} (3/4)(1+c^2) & (-3/4)C s^2 & (3/4)S s^2 & 0 \\ (-3/4)C s^2 & (3/4)(1+c^2)C^2-(3/2)c S^2 & (-3/4)(1+c^2)S C-(3/2)c C S & 0 \\ (-3/4)S s^2 & (3/4)(1+c^2)C S+(3/2)c S C & -(3/4)(1+c^2)S^2+(3/2)c C^2 & 0 \\ 0 & 0 & 0 & (3/2)c \end{vmatrix} \quad (20)$$

Comparison of equation (19) and (20) allows us to assess how dipole scattering is manifested in hemispherical scattering in planetary (normal incidence) and laboratory settings, respectively. It can be seen that laboratory standard settings give more complicated terms (compare $M_{22}$, $M_{23}$, $M_{32}$, $M_{33}$), but the $M_{11}$ term and the last row and column are identical. $M_{21}$ and $M_{31}$ are relatively similar, as is $M_{12}$, however $M_{13}$ is zero in the planetary configuration.

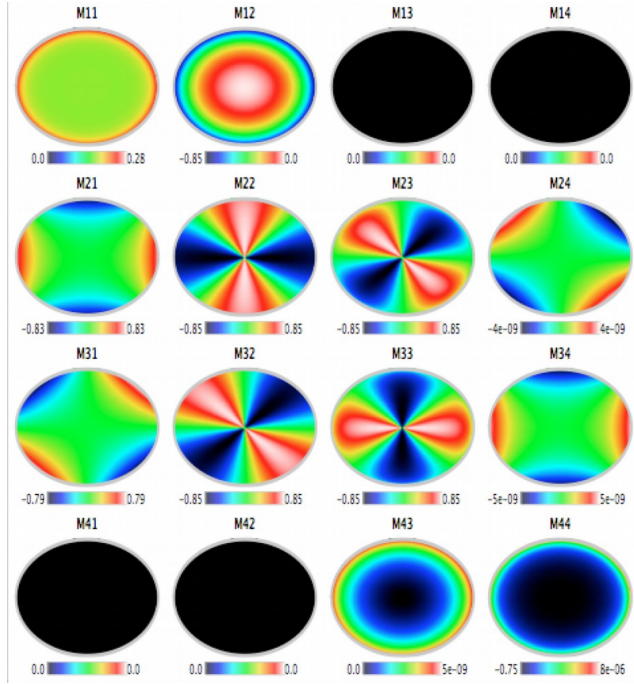

Figure 3 - Müller Matrix images of backscattered photons scattered from a Rayleigh scattering particle. The particle is single scattering and is 0.01 microns in diameter. The laser light is λ=1 micron. The calculations use a spherical target modeled by an adding-doubling approach. The incident light is normal to the target. Color scales indicate intensity relative to the M11 element. See text for discussion of symmetries.

Fig. 3 displays 16 Müller matrix hemispherical maps for a Rayleigh scattering target produced using the adding-doubling approach. This compares favorably with equation (19).

*Notes on Effects of Mirrors in the Beam Path.* The Müller matrix for a perfect mirror for light at normal incidence is [21] (eqn 33.71):

$$M_R = \begin{pmatrix} 1 & 0 & 0 & 0 \\ 0 & 1 & 0 & 0 \\ 0 & 0 & -1 & 0 \\ 0 & 0 & 0 & -1 \end{pmatrix} \quad (21)$$

When the perfect mirror is inclined an angle Θ to the incoming beam, the Müller matrix is modified thus (e.g. MSE Memo [22] 20c, equation (12)):

$$M_R(\theta) = \begin{pmatrix} 1 & 0 & 0 & 0 \\ 0 & \cos 4\theta & \sin 4\theta & 0 \\ 0 & -\sin 4\theta & \cos 4\theta & 0 \\ 0 & 0 & 0 & -1 \end{pmatrix} \quad (22)$$

Therefore when the mirror is inclined 45 degrees to the beam, the ideal mirror Müller matrix becomes:

$$M_R(\theta) = \begin{pmatrix} 1 & 0 & 0 & 0 \\ 0 & -1 & 0 & 0 \\ 0 & 0 & -1 & 0 \\ 0 & 0 & 0 & -1 \end{pmatrix} \quad (23)$$

When we use this scattering matrix in equation for the laboratory/LIDAR scattering geometry, the mirror is inclined 90 degrees to the beam, so the mirror Müller matrix becomes:

$$M_R(\theta) = \begin{pmatrix} 1 & 0 & 0 & 0 \\ 0 & 1 & 0 & 0 \\ 0 & 0 & 1 & 0 \\ 0 & 0 & 0 & -1 \end{pmatrix} \quad (24)$$

**IV. Discussion**

*Müller matrix on and off diagonal symmetries.* Feng [17] noted that the results presented by Hielscher et al. [4] indicated that all off-diagonal Müller matrix maps for Mie and Rayleigh scattering should obey symmetry rules. Feng stated that $M_{12}$ and $M_{21}$, $M_{32}$ and $M_{23}$, and $M_{31}$ and $M_{13}$ should be symmetric, whereas $M_{43}$ and $M_{34}$, $M_{24}$ and $M_{42}$ should be antisymmetric. He established that if instead of eqn (13) one uses R(φ)M'(θ)R(-φ) (his eqn (11)) one obtains these symmetry relationships. He matched this to a range of results in the literature [4,23], and contrasted it with the work of Rakovic et al. [16]. However, Feng did not discuss the relationship between on-diagonal elements $M_{22}$ and $M_{33}$ and their tight linkage to elements $M_{23}$ and $M_{32}$ in the backscattering geometry. When one compares with more recent experimental work [24,25] (not cited by Feng), one realises that it is necessary to adopt the lab phase matrix equation given here (eqn (13)) in order to fulfil the requirement that $M_{22} \sim -M_{33}$ and $M_{23} \sim M_{32}$ (see Figure 2 of [25] and Figure 3 of [24]). The same relationship between $M_{22}$, $M_{33}$, $M_{23}$ and $M_{32}$ is found in Figures 3 and 4 of [23] and Figure 5, 6, 8 and 9 of [4] (which were cited by Feng) so this relationship is likely to be reliable. Simply employing (ideal) mirror reflections, as discussed above, one cannot obtain this symmetry.

Wang et al. [18,26] followed the same modelling approach as Rakovic et al. and their results have similar weaknesses, as pointed out by Feng [17]. In addition, it should be noted that in Rakovic et al., the element $S_{33}$ (in Figure 4 of [16]) is inconsistent with eqn 30b of that same paper.

*Effect of mirror reflections*. It seems likely that the mirror relations (21)-(24) may explain some differences between previous papers in the literature. For example, applying equation (24) on the back and front of Xu's Figure 3 of [24], one obtains a match for Ramella-Roman et al. (Figure 2 of [25]).

In addition, applying eqn (23) back and front to Figure 5, 6, 8 and 9 of Hielscher et al. [4] brings them into concordance with the results of Ramella-Roman et al. (Figure 2 of [25]). Unfortunately, according to the Hielscher et al. description, his optical setup actually involves two 90° mirrors either side of the sample, suggesting application of equation (24) might be more appropriate. We leave resolution of this question for future work.

We have followed a strategy of trusting the experimental results reported in Ramella-Roman et al. [25] because these are the most recent laboratory results from an experienced optical group. However, the mathematical approach in Ramella-Roman was not fully characterized, because simple descriptions of an optical setup were given with no mathematical relationship to govern the effect of the mirrors mentioned (in the text on p. 10394 of [25]). Undoubtedly, the results presented here are not the last word on this matter, but one can hope that the direction of future research is now made somewhat clearer. Non ideal mirrors may also play a role in the final resolution of these relationships. Further experimental work is needed to verify the relations (13) in the laboratory with a well characterised optical configuration, including knowledge of mirror angles and non-ideal properties.

*Applications and comparisons to other recent Müller matrix studies*. Müller matrix analysis has recently been used for size determination of industrial products [27,28].

Biomedical applications of Müller matrix measurements have sought to use polarimetry to image cancerous cells through tissue [29]. To test and calibrate the data gathered from these biomedical imagers, the Monte Carlo schemes we have discussed are very popular [5,23,30,31]. Hopefully the results of this paper will be helpful in this area.

*Aniostopic scatterers*. In a very interesting recent study, Alali et al. [32] used a Monte Carlo approach to examine the axial heterogenity of birefringent materials using polarized light. They developed an "asymmetry degree (ASD)" in eqns 18 and 19 based on 6 off diagonal elements of the Müller matrix and tested the ASD metric experimentally on stretched polyamide samples which provide an anisotropic experimental target. In addition, Zeng et al. [33] developed rotation independent polarization Müller matrix based parameters for use in characterization of anisotropic scattering objects, which require a full 16-element Müller matrix because they are anisotropic scatterers. Sun et al. [34] then tested their applicability on textile samples.

Ossikovski et al. have developed several alternative decompositions of the Müller matrix and have addressed the symmetries of these decompositions [35,36]. In addition, Gil and Jose have recently presented an algebraic decomposition of Müller matrices [37].

Yurkin [38] approached the polarization measurement problem from several different angles. His approach was to look at integrals over the azimuth, in order to simulate the Müller matrix symmetries measured using flow cytometry.

## V. Conclusions

This paper has outlined the theoretical linkages between the scattering Müller matrices for "planetary" and "LIDAR/laboratory" observations. Thus, the scene is now set for further advances linking the fields of planetary and laboratory or field observations. Instrument development [39] to study these linkages should go hand in hand with the theories developed here to investigate aspects such as curve fitting of spectral observations of polarization arising from absorption bands [40], broken symmetries and intercomparisons of Monte Carlo and adding-doubling methods. The successful application analysis of data for terrestrial and planetary missions (such as recently proposed by [41]) utilizing polarization observations relies on interpretations such as that carried out here.

## VI. Note on arXiv preprint service

This paper will be submitted to the free arXiv, and as errors are reported to the author in the mathematical formulas herein, the arXiv copy will be kept up to date.

## VII. Acknowledgements

This work was partly supported by two grants (NNX11AP23G and NNX13AN21G) from the NASA Planetary Geology and Geophysics program run by Dr. Mike Kelley. I also thank the organizers of the arXiv pre-print service.